\documentclass[aps,twocolumn,floats,prl]{revtex4}
\usepackage{graphics,graphicx,epsfig}
\usepackage{amssymb,color}
\usepackage{epsf,epstopdf,wrapfig}

\begin{document}

\title{The emergence of stereotyped behaviors in {\em C. elegans}}

\author{Greg J Stephens, William S Ryu\footnote{Present address: Department of Physics, Banting and Best Department of Medical Research,University of Toronto, Toronto M5S 1A7, Ontario, Canada} and William Bialek}

\affiliation{Joseph Henry Laboratories of Physics and
Lewis--Sigler Institute for Integrative Genomics,
Princeton University, Princeton, New Jersey 08544 USA}

\begin{abstract}
Animal behaviors are  sometimes decomposable into discrete, stereotyped elements.  In one model, such behaviors are triggered by specific commands; in the extreme case, the discreteness of behavior is traced to the discreteness of action potentials in the individual command neurons.  We use the crawling behavior of the nematode {\em C. elegans}  to explore the opposite extreme, in which discreteness and stereotypy emerges from the dynamics of the entire behavior.   A simple stochastic model for the worm's continuously changing body shape during crawling has attractors corresponding to forward and backward motion; noise--driven transitions between these attractors correspond to abrupt reversals.  We show that, with no free parameters, this model generates reversals at a rate within error bars of that observed experimentally, and the relatively stereotyped trajectories in the neighborhood of the reversal also are predicted correctly.
\end{abstract}

\date{\today}

\maketitle

Many organisms, from bacteria to humans, exhibit discrete, stereotyped motor behaviors.  A common model is that these behaviors are stereotyped because they are triggered by specific commands, and in some cases we can identify ``command neurons'' whose activity provides the trigger \cite{bullock}.  In the extreme, discreteness and stereotypy of the behavior reduces to the discreteness and stereotypy of the action potentials generated by the command neurons, as with the escape behaviors in fish triggered by spiking of the Mauthner cell \cite{mauthner}.  But the stereotypy of spikes itself emerges from the continuous dynamics of currents, voltages and ion channel populations \cite{hodgkin+huxley_52d,modHH}.  Is it possible that, in more complex systems, stereotypy emerges not form the dynamics of single neurons, but from the dynamics of larger circuits of neurons, perhaps coupled to the mechanics of the behavior itself?  Here we explore this possibility in the context of crawling behavior in the nematode {\em C. elegans} \cite{croll_75}.   These worms generate abrupt reversals of direction \cite{zhao+al_03,gray+al_05}, and it is the emergence of these discrete events that we try to understand.  

The problem of reversals in {\em C. elegans} is interesting in part because the underlying neural circuitry includes a nominal command neuron, AVA \cite{chalfie+al_85}, whose activity is correlated with forward vs. backward crawling \cite{chronis+al_07}.   On the other hand, AVA is an interneuron in a network, and it is not clear whether the decision to reverse direction can be traced to a single cell.  Even when AVA is ablated, reversals occur, although the distribution of times spent in the backward crawling state shifts \cite{gray+al_05}. Further, most neurons in {\em C. elegans} are thought not to generate action potentials, so even if a single neuron dominates the decision it is not obvious why the trajectory of a reversal would be stereotyped.  
Rather than probing further into the neural circuitry, it may be useful to step back and give a more quantitative description of the reversal behavior itself.

\begin{figure}[b]
\begin{center}
\includegraphics[width=0.95\columnwidth]{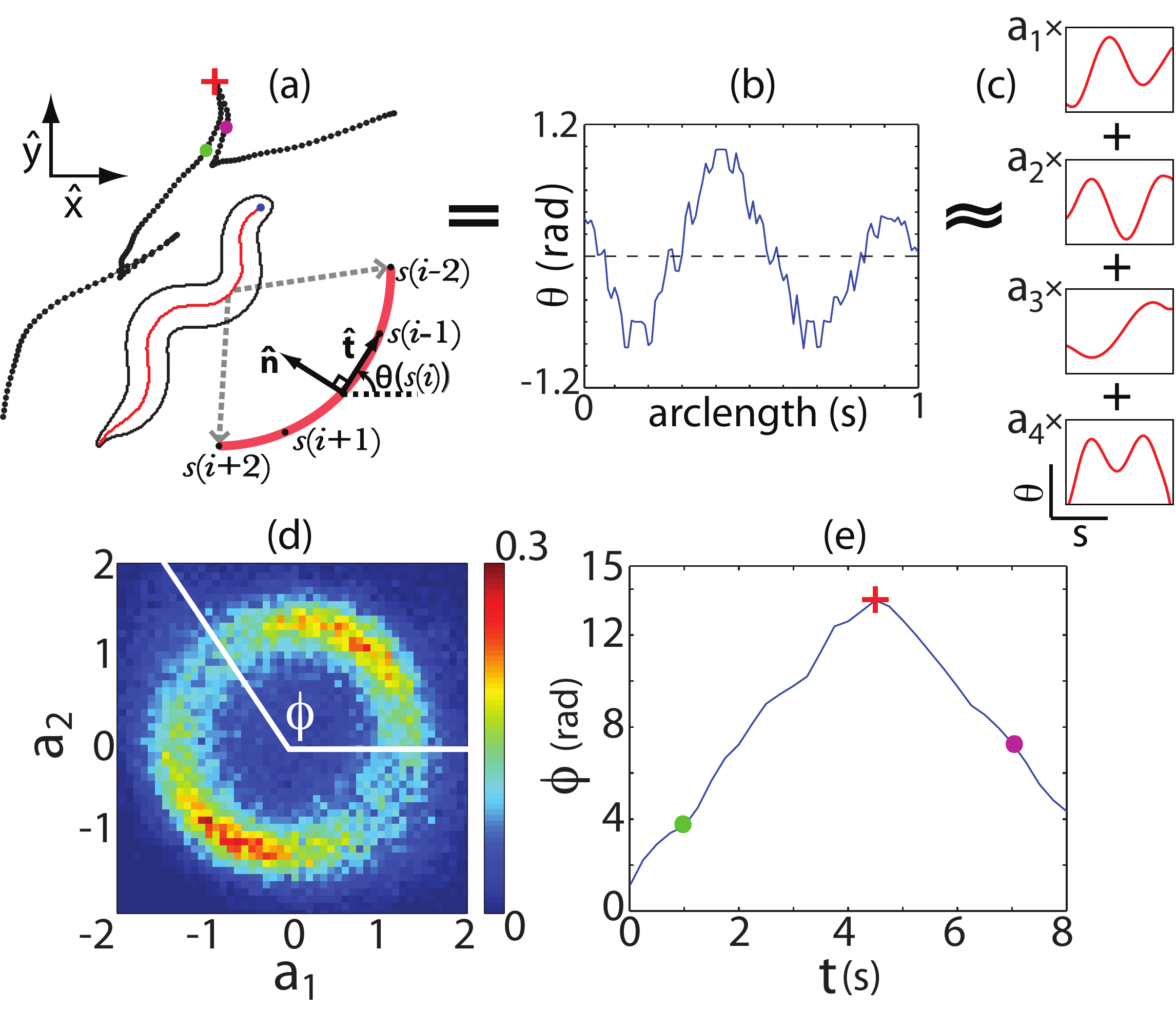}
\caption{Reversals in shape space, following Ref \cite{stephens+al_08}.  (a) Tracking video microscopy gives both the x--y trajectory of the worm as it crawls on an agar plate, and the shape of the worm's body at high resolution.
(b)  Shape is described by the tangent angle $\theta$ vs. arc length $s$, in intrinsic coordinates such that $\int ds\, \theta (s) = 0$.
(c)  We decompose $\theta(s)$ into four dominant modes.
(d) Amplitudes along the first two modes oscillate, with nearly constant amplitude but time varying phase $\phi = \tan^{-1}(a_2 / a_1)$; here the amplitudes are normalized so that $\langle a_{\rm i}^2\rangle =1$.
(e)  The phase trajectories exhibit abrupt reversals, moments when $\omega \equiv d\phi/dt$ change sign.  Comparing the points marked here and in (a), these phase reversals correspond to reversals in crawling direction.
\label{fig1}}
\end{center}
\end{figure}

Locomotion involves changes of body shape.  In previous work \cite{stephens+al_08}, we have have shown that as {\em C. elegans} crawls, its body moves through a ``shape space'' of restricted dimensionality.  Of the four dimensions that capture $\sim 95\%$ of the variance in body shape, oscillatory motions along the first two modes correspond to the propulsive wave which passes along the worm's body and drives it forward or backward.  Indeed, the rate at which the phase of this oscillation changes predicts, quantitatively, the velocity of the worm's motion \cite{stephens+al_09a}.  As emphasized in Fig \ref{fig1}, this correspondence includes the fact that abrupt changes in the sign of the phase velocity predict the  points where the worm suddenly ``backs up'' and reverses its crawling direction.   

\begin{figure}[b]
\begin{center}
\includegraphics[width=0.9\columnwidth]{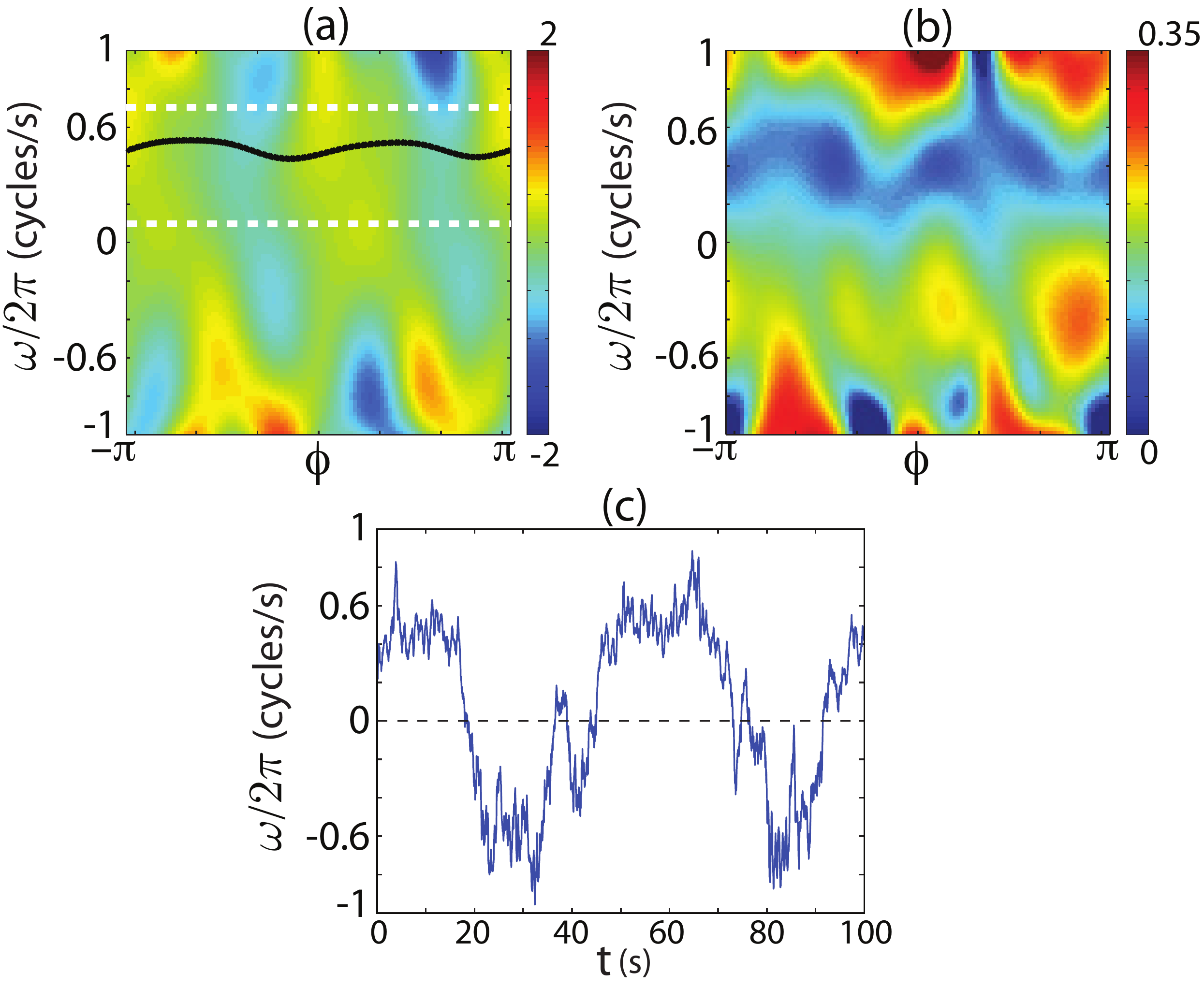}
\caption{Langevin model for the phase dynamics,  Eq's (\ref{def_omega}, \ref{langevin_omega}). 
(a) The deterministic component of the force $F(\omega, \phi)$, in units ${\rm cycles/s}^2$.  The black line is the attracting limit cycle corresponding to forward crawling, and the white dashed lines mark boundaries for our analysis of trajectories that start within this attractor.
(b) The state dependent noise $\sigma(\omega,\phi)$, in units ${\rm cycles/s}^{3/2}$.    (c) A sample of the trajectories resulting from  Eq's (\ref{def_omega}, \ref{langevin_omega}), illustrating  transitions between attractors at positive and negative $\omega$, corresponding to forward and backward crawling.\label{langevin_fig}}
\end{center}
\end{figure}

In Ref \cite{stephens+al_08} we constructed a simple model for the dynamics of the phase $\phi(t)$.  Since the worm can crawl both forward and backward, a minimal for the phase dynamics is a second order system.  Since the dynamics are noisy, we try to write something analogous to the Langevin equation for a Brownian particle subject to forces.  We recall that for the Brownian particle, we have
\begin{eqnarray}
{dx  \over dt}&=& v  ,  \label{langevin1}\\
m{dv \over dt}&=&f(x,v) +\xi(t), 
\label{langevin2}
\end{eqnarray}
where $m$ is the mass of the particle, $f(x,v)$ describes the average forces acting on the particle, and $\xi(t)$ is the random force resulting from molecular collisions.  By analogy, then, we write for the phase of the worm's shape oscillations
\begin{eqnarray}
{d\phi  \over dt}&=& \omega  ,  \label{def_omega}\\
{d\omega \over dt}&=&F(\omega,\phi) +\sigma(\omega,\phi)\eta(t).  
\label{langevin_omega}
\end{eqnarray}
Here we allow the possibility that, unlike a Brownian particle in equilibrium at a fixed temperature, the strength of the noise varies with the state of the system.
Will still assume, however, that the noise reflects events on very short time scales, so that
\begin{equation}
\langle \eta(t) \eta (t') \rangle = \delta (t-t').
\end{equation}

There is a substantial literature on how one learns  Langevin models from real data; see, for example, Refs \cite{langevinEOM1,langevinEOM2}.  A central difficulty is not to overfit by allowing for arbitrarily complex functions describing the force.  To regularize the learning problem we assumed that the force could be written as a polynomial in $\omega$ and a Fourier series in $\phi$,
\begin{equation}
F(\omega,\phi ) = \sum_{p=0}^{P} \sum_{m=-M}^M \alpha_{mp} e^{-im\phi} \omega^p .
\end{equation}
Then if we have long trajectories $\phi(t)$, the parameters $\alpha_{mp}$ are those which minimize
\begin{equation}
\chi^2 = {\bigg\langle}
\left[ \ {{d\omega} \over {dt}} - F(\omega , \phi)\right]^2 
{\bigg\rangle},
\end{equation}
where the average is computed over the long trajectory; in practice we average over five long trajectories from each of twelve worms, for a total of two hours of data.  The optimal choice of the series orders $P$ and $M$ are found by fitting to $90\%$ of the data and minimizing the generalization error computed on the remaining $10\%$; we find $P = M = 5$.  Note also that the trajectories are given experimentally as discrete time samples, here with time steps $\Delta t = 1/32 \,{\rm s}$, so that all time derivatives must be constructed carefully; to minimize the impact of measurements errors we smooth the mode amplitudes $a_1(t)$ and $a_2(t)$ with fourth order polynomials before computing the phase.    Finally, the noise strength is defined by
\begin{equation}
\sigma^2 (\omega , \phi ) = \Delta t {\bigg\langle}
\left[ \ {{d\omega} \over {dt}} - F(\omega , \phi)\right]^2 
{\bigg\rangle}_{\omega , \phi},
\end{equation}
where the average now is taken over those moments in the data when the state of the system is characterized by particular values of $\omega$ and $\phi$.  The results of this construction are shown in Fig \ref{langevin_fig}a and b \cite{stephens+al_08}.

It is important to emphasize that the construction of the Langevin model allows us to look only at local features of the phase trajectory; we do not use, directly, any information about what happens on long time scales.  Nonetheless, as in the corresponding physics problems, the model predicts a variety of phenomena that emerge on long time scales.  As described in Ref \cite{stephens+al_08}, the underlying deterministic model (where we set $\sigma = 0$) has multiple attractors, limit cycles corresponding to forward and  backward crawling, and fixed points corresponding to pauses.  In the full dynamics with noise, the system is predicted to remain near these attractors for extended periods of time.  The noise drives random motions in the neighborhood of the attractors, as well as phase diffusion along the limit cycles; these are effects that we can think of as perturbations to the deterministic dynamics.  There is also a non--perturbative effect:  noise drives sudden transitions from one attractor to another, as seen in Fig \ref{langevin_fig}c.  In particular,   there are transitions from the $\omega >0$ attractor to the $\omega < 0$ attractor, and these should correspond to reversals in the direction of crawling, as seen in Fig \ref{fig1}a.

\begin{figure}[t]
\begin{center}
\includegraphics[width=\columnwidth]{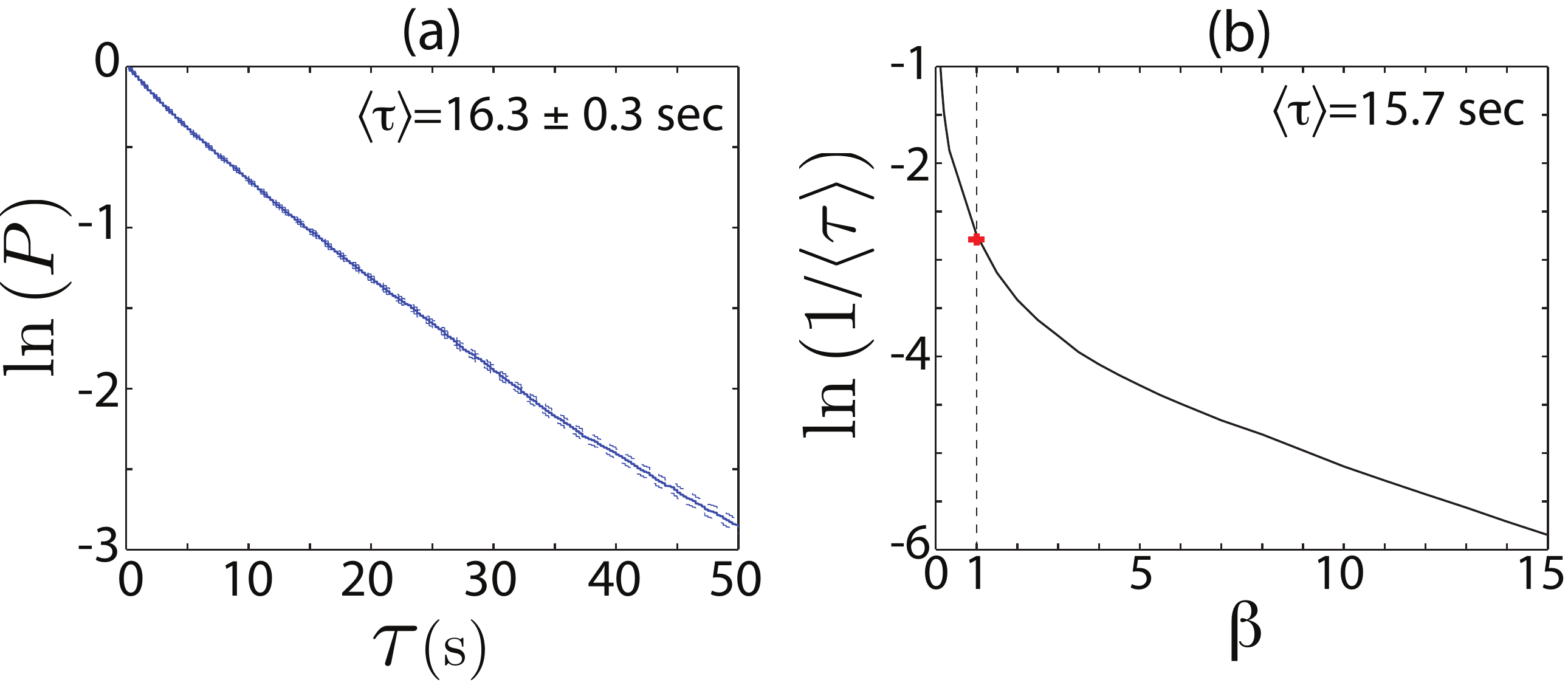}
\caption{Survival times.  (a) We measure the probability that a worm's trajectory which is in the neighborhood of the froward attractor at time $t$ has not crossed to negative phase velocity by time $t+\tau$.  The decay is exponential, with a mean time $\langle\tau\rangle = 16.3\pm 0.3\,{\rm s}$.
(b)  The predicted mean time $\langle\tau\rangle$ depends on the noise level in the model.  If we scale $\sigma^2$ by a factor $\beta$, then for large $\beta$ we find $1/\langle\tau\rangle \propto \exp(-\beta E)$, analogous to the Arrhenius temperature dependence of chemical reaction rates.  Cross shows the measured $1/\langle\tau\rangle$ at the real temperature $\beta = 1$; vertical error bar is the standard deviation of the reversal rate, and horizontal error bar is the standard deviation of the noise level, both computed across the population of worms. 
\label{intervals}}
\end{center}
\end{figure}

To quantify the predicted and observed reversals, we measure the survival probability in the forward crawling attractor.  As we observe the trajectory $\phi(t)$, we choose, at random, a moment in time where phase velocity $0.1 < \omega < 0.6\,{\rm rad/s}$, a region indicated by the dashed white lines in Fig \ref{langevin_fig}a.  Then we declare that a reversal has occurred if the phase velocity falls below zero; the survival probability $P(\tau)$ is the probability that this has {\em not} happened after a delay $\tau$.  If transitions are the result of brief events, well separated in time, then there should be no memory form one to the next, and we expect the survival probability  to decay exponentially, $P(\tau ) = \exp(-\tau/\langle\tau\rangle )$; this is what we observe both in simulated trajectories and in the actual data, as shown in Fig \ref{intervals}.  In the data, the mean interval is $\langle\tau\rangle = 16.3\pm 0.3\,{\rm s}$, where the error is the standard deviation across 33 worms, each observed for 35 minutes; this is a completely independent data set, with different individual worms, from that used in learning the Langevin model.  The model predicts $\langle\tau\rangle = 15.7\,{\rm s}$, which agrees within $4\%$ accuracy.

The escape from one attractor to another under the influence of noise is like the escape from one metastable configuration to another via Brownian motion---a chemical reaction \cite{hanggi+al_90}.  The strength of the noise, $\sigma^2$ plays the role of temperature, and we expect that if the temperature changes we should see the Arrhenius law, as shown in Fig \ref{intervals}b.  The actual ``temperature'' is a bit too high for the Arrhenius law to be valid, but the mean time between intervals is still an order of magnitude longer than the characteristics times for motion within the forward or backward crawling attractors.    Also, when we estimate the noise level from the trajectories, there is an error in our estimate, and this propagates to give an error in the predicted mean time between attractors which is comparable to the deviation between the prediction and the data.  We conclude that noise--driven escape from the forward crawling attractor provides a quantitatively accurate model for the observed rate of reversals, without introducing any new parameters; the long time between reversals emerges from the dynamics in the same way the long time between chemical reaction events emerges from the fast Brownian dynamics of the molecules.

\begin{figure}[b]
\begin{center}
\includegraphics[width=\columnwidth]{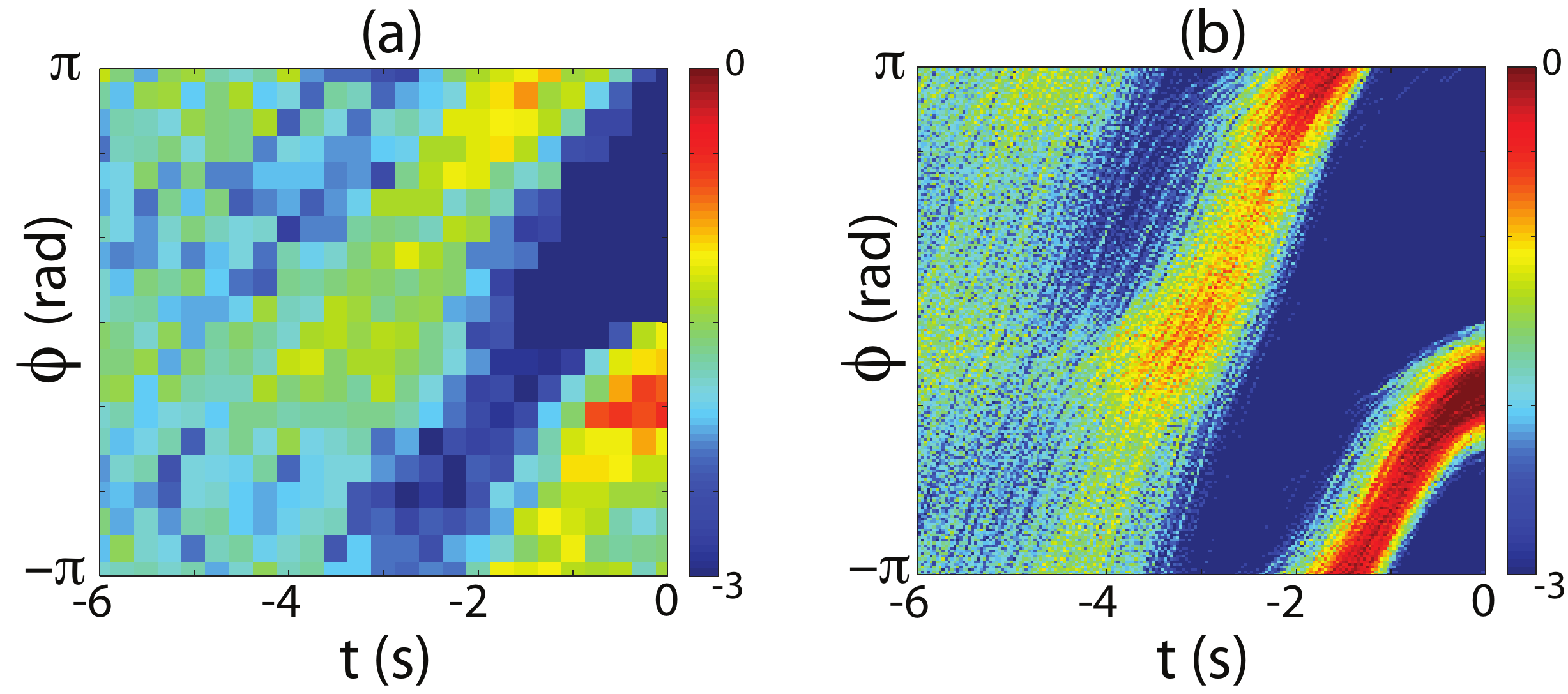}
\caption{Emergence of stereotyped behaviors. (a) The conditional density $\rho(\phi | t)$ constructed
from an ensemble of worm trajectories aligned to exit the forward attractor at $t=0$ via a path with $\phi(0) < 0$.  Color scale is for $\ln [\rho \cdot (1\, {\rm rad})]$.  (b) The same
density generated from simulations of the stochastic model, Eq's (\ref{def_omega}, \ref{langevin_omega}). 
\label{stereotypy}}
\end{center}
\end{figure}

One of the important results in the theory of thermally activated escape over a barrier is that, in the low noise limit, the escape trajectories become stereotyped \cite{dykman+al_94,notes}.    By analogy, then, we expect that the trajectories that allow the worm to escape from the forward crawling attractor also should be stereotyped, or at least clustered around some prototypical trajectories.  Detailed analysis of the simulations show that there are in fact two such clusters, corresponding to transitions in which the sign of $\omega$ changes while the phase $\phi$ is positive or negative; this doubling is also seen in the data (not shown).  If we focus on the transitions that occur with negative phase, we can align all the phase trajectories at the moment where $\omega$ changes sign, and estimate the probability distribution $\rho(\phi|t)$ at any time $t$ relative to the switch.  As we see in Fig \ref{stereotypy}, both the real data and the simulations show that this distribution is concentrated fairly tightly, and this extends back for several seconds before the moment of the reversal itself.  Comparing Figs \ref{stereotypy}a and b,  we see that the conditional density derived from worm motion appears as a noisy version of the density derived from the stochastic dynamics.

In summary, we have found a surprisingly simple model for the dynamics of {\em C. elegans} crawling.    The construction of the model depends on analyzing trajectories over very short time scales, essentially trying to map the way in which acceleration depends on position and velocity in a simple phase space.  But if we take this local description seriously, we predict phenomena on much longer time scales.  As with models of single neurons and small circuits, our model has multiple attractors, which we can identify with distinct behavioral states, and spontaneous transitions among these attractors.  It is one class of these spontaneous transitions, reversals, which we have focused on here.  Because the transitions are driven by noise, the rate of transitions is suppressed exponentially relative to the natural time scales of the dynamics, in the same way that chemical reaction rates are exponentially slower than the time scales  of small amplitude molecular motions.    We find that the reversal rates predicted by the model agree with experiment with an accuracy of $4\%$, within the errors of our estimates of the underlying noise levels.  In more detail, the model predicts that the reversals occur via stereotyped trajectories, and these too agree with experiment.   Rather than being traced to discrete, stereotyped commands, the stereotypy of reversals is an emergent property of the behavioral dynamics as a whole.

\acknowledgments{We thank T Mora,  S Norrelykke and G Tka\v{c}ik for discussions, and B Johnson--Kerner for help with the original experiments on which this analysis is based.
This work was supported in part by NIH grants P50 GM071508 and R01 EY017210, by NSF grant  PHY--0650617,  and by the Swartz Foundation.}

\end{document}